# Effective Computation of Immersion Obstructions for Unions of Graph Classes

Archontia C. Giannopoulou[*][†]   Iosif Salem[*]   Dimitris Zoros[*]

**Abstract**

In the final paper of the Graph Minors series N. Robertson and P. Seymour proved that graphs are well-quasi-ordered under the immersion ordering. A direct implication of this theorem is that each class of graphs that is closed under taking immersions can be fully characterized by forbidding a finite set of graphs (immersion obstruction set). However, as the proof of the well-quasi-ordering theorem is non-constructive, there is no generic procedure for computing such a set. Moreover, it remains an open issue to identify for which immersion-closed graph classes the computation of those sets can become effective. By adapting the tools that were introduced by I. Adler, M. Grohe and S. Kreutzer, for the effective computation of minor obstruction sets, we expand the horizon of computability to immersion obstruction sets. In particular, our results propagate the computability of immersion obstruction sets of immersion-closed graph classes to immersion obstruction sets of finite unions of immersion closed graph classes.

**Keywords:** Immersions, Obstructions, Unique Linkage Theorem, Treewidth

## 1 Introduction

The development of the graph minor theory constitutes a vital part of modern Combinatorics. A lot of theorems that were proved and techniques that were introduced in its context, appear to be of crucial importance in Algorithmics and the theory of Parameterized Complexity as well as in Structural Graph Theory. Such examples are the Excluded Grid Theorem [30], the

[*]Department of Mathematics, National and Kapodistrian University of Athens. Emails: {arcgian,ysalem,dzoros}@math.uoa.gr

[†]The first author was supported by a grant of the Special Account for Research Grants of the National and Kapodistrian University of Athens (project code: 70/4/10311) and part of the work of this author has received funding from the European Research Council under the European Union's Seventh Framework Programme (FP7/2007-2013)/ERC grant agreement no. 259385.

Structural Theorems in [28, 31] and the Irrelevant Vertex Technique in [27]. (For examples of algorithmic applications, see [12, 22]).

We say that a graph $H$ is an immersion (minor) of a graph $G$, if we can obtain $H$ from a subgraph of $G$ by lifting (contracting) edges. (For detailed definitions, see Section 2). While the minor ordering has been extensively studied throughout the last decades [1, 8, 27, 28, 30–33], the immersion ordering has only recently gained more attention [13, 22, 34]. One of the fundamental results that appeared in the last paper of the Graph Minors series was the proof of Nash-Williams' Conjecture, that is, the class of all graphs is well-quasi-ordered by the immersion ordering [33]. A direct corollary of these results is that a graph class $\mathcal{C}$, which is closed under taking immersions, can be characterized by a finite family $\mathbf{obs}_{\leq_{im}}(\mathcal{C})$ of minimal, according to the immersion ordering, graphs that are not contained in $\mathcal{C}$ (called obstructions from now on). Furthermore, in [22], it was proven that there is an $O(|V(G)|^3)$ algorithm that decides whether a graph $H$ is an immersion of a graph $G$ (where the hidden constants depend only on $H$). Thus, an immediate algorithmic implication of the finiteness of $\mathbf{obs}_{\leq_{im}}(\mathcal{C})$ and the algorithm in [22], is that it can be decided in cubic time whether a graph belongs to $\mathcal{C}$ or not (by testing if the graph $G$ contains any of the graphs in $\mathbf{obs}_{\leq_{im}}(\mathcal{C})$ as an immersion). In other words, these two results imply that membership in an immersion-closed graph class can be decided in cubic time.

We would like to mention here that the same meta-algorithmic conclusion holds for the minor ordering from the proofs in [28] and [29]. Evenmore, this result, that is, the existence of a cubic time algorithm deciding the membership of a graph in a graph class that is closed under minors, broadened the perspectives towards the understanding of the NP-hard problems. It was actually at that point that it became clear what seemed to be as "*different levels of hardness*" between these problems [5]. Notice for example, for the well-known $k$-VERTEX COVER problem, that the class of graphs admitting a vertex cover of size at most $k$ is closed under taking minors. Therefore, for every fixed $k$ there is a cubic time algorithm deciding whether a graph has a vertex cover of size $k$. However, no similar result can be expected for the $k$-COLORING problem, as it is known to be NP-hard for every fixed $k \geq 3$. The observation of this gap in the time complexity of the NP-hard problems facilitated the development of the Parameterized Complexity Theory [14, 19, 26] by M. Fellows and R. Downey, which has proven to be a very powerful theory and has majorly advanced during the past decades (for example, see [2, 6, 7, 11, 12]).

Nevertheless, the aforementioned meta-algorithmic result for an immersion-closed graph class $\mathcal{C}$ assumes that the family $\mathbf{obs}_{\leq_{im}}(\mathcal{C})$ is known. Evenmore, as the proofs in [29] and [33] are non-constructive (see [20]), no generic algorithm is provided that allows us to identify these obstruction sets for every immersion-closed graph class. Moreover, even for fixed graph classes, this task can be extremely challenging as such a set could contain many graphs and no general upper bound on its cardinality is known other than its finiteness [15]. The issue of the computability of obstruction sets for minors and immersions was raised by M. Fellows and M. Langston [17, 18] and the challenges against computing obstruction sets soon became clear. In particular, in [18] M. Fellows and M. Langston showed that the problem of determining obstruction sets from machine descriptions of minor-closed graph classes is recursively unsolvable (which directly holds for the immersion ordering as well). Evenmore, in [10] B. Courcelle, R. Downey and M. Fellows proved that the obstruction set of a minor-closed graph class cannot be computed from a description of the minor-closed graph class in Monadic Second Order Logic (MSO). Thus, a consequent open problem is to identify the information that is needed for an immersion-closed graph class $\mathcal{C}$ in order to make it possible to effectively compute the obstruction set $\mathbf{obs}_{\leq_{im}}(\mathcal{C})$.

Several methods have been proposed towards tackling the non constructiveness of these sets (see, for example, [8, 17]) and the problem of algorithmically identifying minor obstruction sets has been extensively studied [1, 8, 10, 17, 18, 24]. In [8], it was proven that the obstruction set of a minor-closed graph class $\mathcal{F}$ which is the union of two minor-closed graph classes $\mathcal{F}_1$ and $\mathcal{F}_2$ whose obstruction sets are given can be computed under the assumption that there is at least one tree that does not belong to $\mathcal{F}_1 \cap \mathcal{F}_2$ and in [1] it was shown that the aforementioned assumption is not necessary.

In this paper, we initiate the study for computing immersion obstruction sets. In particular, we deal with the problem of computing $\mathbf{obs}_{\leq_{im}}(\mathcal{C})$ for families of graph classes $\mathcal{C}$ that are constructed by finite unions of immersion-closed graph classes. Notice that the union and the intersection of two immersion-closed graph classes are also immersion-closed, hence their obstruction sets are of finite size. It is also easy to see that, given the obstruction sets of two immersion-closed graph classes, the obstruction set of their intersection can be computed in a trivial way. We prove that there is an algorithm that, given the obstruction sets of two immersion-closed graph classes, outputs the obstruction set of their union.

Our approach is based on the derivation of an upper bound on the tree-width of the obstructions of an immersion-closed graph class. Notice that the combination of a machine description of an immersion-closed graph class $\mathcal{F}$ with an upper bound on the size of the forbidden graphs makes this computation possible, but neither the machine description of the class nor the upper bound alone are sufficient information. Moreover, as mentioned before, no generic procedure is known for computing such an upper bound. We build on the machinery introduced by I. Adler, M. Grohe and S. Kreutzer in [1] for computing minor obstruction sets. In particular, we will ask for an MSO-description of an immersion-closed graph class instead of a machine description, and a bound on the tree-width instead of an upper bound on the size of the obstructions of the immersion-closed graph class.

For this, we adapt the results on [1] so to permit the computation of the obstruction set of any immersion-closed graph class, under the conditions that an explicit upper bound on the tree-width of its obstructions can also be computed and the class can be defined in MSO. We present this algorithm at Lemma 4, and with that we conclude the computability part of the paper. Our next step is a combinatorial result proving an upper bound on the tree-width of the obstructions of the union of two immersion-closed graph classes, whose obstruction sets are known. We then show that the obstruction set of their union can be effectively computed. Our combinatorial proofs significantly differ from the ones in [1] and make use of a suitable extension of the Unique Linkage Theorem of K. Kawarabayashi and P. Wollan [23].

The remainder of the paper is structured as follows. In Section 2 we state the basic notions that we use throughout the paper as well as few well-known results. In Section 3 we present our computability result, that is, we prove that the obstruction set of an immersion-closed graph class can be computed when an upper bound on the tree-width of its obstructions and an MSO-description of the graph class are known. We do so by proving a version of Lemma 3.1 of [1], adapted to the immersion ordering. In Section 4 we provide the bounds on the tree-width of the graphs that belong in $\mathbf{obs}_{\leq_{im}}(\mathcal{C}_1 \cup \mathcal{C}_2)$ by assuming that the sets $\mathbf{obs}_{\leq_{im}}(\mathcal{C}_1)$ and $\mathbf{obs}_{\leq_{im}}(\mathcal{C}_2)$ are known, where $\mathcal{C}_1$ and $\mathcal{C}_2$ are immersion-closed graph classes. By doing this we propagate the computability of immersion obstruction sets to finite unions of immersion-closed graph classes.

## 2 Preliminaries

### 2.1 Basics

Throughout this paper, graphs are unweighted, undirected and contain no loops or multiple edges unless otherwise specified. Given a graph $G$, we denote its set of vertices with $V(G)$, its set of edges with $E(G)$ and the *degree* of a vertex $v$ with $\deg_G(v)$. The *line graph* of a graph $G$, denoted by $L(G)$, is the graph $(E(G), X)$, where $X = \{\{e_1, e_2\} \subseteq E(G) \mid e_1 \cap e_2 \neq \emptyset \wedge e_1 \neq e_2\}$. Given two graphs $G$ and $H$, the *lexicographic product* $G \times H$, is the graph with $V(G \times H) = V(G) \times V(H)$ and $E(G \times H) = \{\{(x,y),(x',y')\} \mid (\{x,x'\} \in E(G)) \vee (x = x' \wedge \{y,y'\} \in E(H))\}$.

Given an edge $e = \{x, y\}$ of a graph $G$, the graph $G/e$ is obtained from $G$ by contracting the edge $e$, that is, the endpoints $x$ and $y$ are replaced by a new vertex $v_{xy}$ which is adjacent to the old neighbors of $x$ and $y$ (except $x$ and $y$). A graph $H$ is a *minor* of $G$, $H \leq_m G$, if there is a function that maps every vertex $v$ of $H$ to a connected set $B_v \subseteq V(G)$, such that for every two distinct vertices $v, w$ of $H$, $B_v$ and $B_w$ share no common vertex, and for every edge $\{u, v\}$ of $H$, there is an edge in $G$ with one endpoint in $B_v$ and one in $B_u$. The graph that is obtained by the union of all $B_v$ such that $v \in V(H)$ and by the edges between $B_v$ and $B_u$ in G, if there exists an edge $\{v, u\}$ in $H$, is called a *model of H in G*. A model with minimal number of vertices and edges is called *minimal model*.

We say that $H$ is an *immersion* of $G$ (or $H$ is *immersed* in $G$), $H \leq_{im} G$, if $H$ can be obtained from a subgraph of $G$ after a (possibly empty) sequence of edge lifts, where the *lift* of two edges $e_1 = \{x, y\}$ and $e_2 = \{x, z\}$ to an edge $e$ is the operation of removing $e_1$ and $e_2$ from $G$ and then adding the edge $e = \{y, z\}$ in the resulting graph (here is the only exception where the existence of multiple edges and loops is allowed). Equivalently, we say that $H$ is an immersion of $G$ if there is an injective mapping $f : V(H) \to V(G)$ such that, for every edge $\{u, v\}$ of $H$, there is a path from $f(u)$ to $f(v)$ in $G$ and for any two distinct edges of $H$ the corresponding paths in $G$ are *edge-disjoint*, that is, they do not share common edges. Additionally, if these paths are internally disjoint from $f(V(H))$, then we say that $H$ is *strongly immersed* in $G$. As above, the function $f$ is called a *model of H in G* and a model with minimal number of vertices and edges is called *minimal model*. A graph class $\mathcal{C}$ is called *immersion-closed*, if for every $G \in \mathcal{C}$ and every $H$ with $H \leq_{im} G$ it holds that $H \in \mathcal{C}$. For example, the class of graphs $\mathcal{E}_t$ that admit a proper edge-coloring of at most $t$ colors such that for every two edges of the same color every path between them contains an edge of greater color is immersion closed. (See [4]). Two paths are called *vertex-disjoint* if they do not share common vertices.

We define an ordering $\leq$ between finite sets of graphs as follows: $\mathcal{F}_1 \leq \mathcal{F}_2$ if and only if

1. $\sum_{G \in \mathcal{F}_1} |V(G)| < \sum_{H \in \mathcal{F}_2} |V(H)|$ or

2. $\sum_{G \in \mathcal{F}_1} |V(G)| = \sum_{H \in \mathcal{F}_2} |V(H)|$ and $\sum_{G \in \mathcal{F}_1} |E(G)| < \sum_{H \in \mathcal{F}_2} |E(H)|$.

**Definition 1.** *Let $\mathcal{C}$ be an immersion-closed graph class. A set of graphs $F = \{H_1, \ldots, H_n\}$ is called (immersion) obstruction set of $\mathcal{C}$, and is denoted by $\mathbf{obs}_{\leq_{im}}(\mathcal{C})$, if and only if $F$ is a $\leq$-minimal set of graphs for which the following holds: For every graph $G$, $G$ does not belong to $\mathcal{C}$ if and only if there exists a graph $H \in F$ such that $H \leq_{im} G$.*

*Remark* 1. We would like to remark here that the obstruction set of an immersion-closed graph class can equivalently be defined in the following way: For any immersion-closed graph class $\mathcal{C}$, the set of its obstructions is the set consisting of all $\leq_{im}$-minimal elements that do not belong in $\mathcal{C}$. However, we also include Definition 1 as it may facilitate the understanding of the intuition behind Lemma 4.

Recall that, because of the seminal result of N. Robertson and P. Seymour [33], for every immersion-closed graph class $\mathcal{C}$, the set $\mathbf{obs}_{\leq_{im}}(\mathcal{C})$ is finite.

## 2.2 Tree-width and Linkages

A *tree decomposition* of a graph $G$ is a pair $(T, B)$, where $T$ is a tree and $B$ is a function that maps every vertex $v \in V(T)$ to a subset $B_v$ of $V(G)$ such that:

(i) for every edge $e$ of $G$ there exists a vertex $t$ in $T$ such that $e \subseteq B_t$,

(ii) for every $v \in V(G)$, if $r, s \in V(T)$ and $v \in B_r \cap B_s$, then for every vertex $t$ on the unique path between $r$ and $s$ in $T$, $v \in B_t$ and

(iii) $\cup_{v \in V(T)} B_v = V(G)$.

The *width of a tree decomposition* $(T, B)$ is $\mathrm{width}(T, B) := \max\{|B_v| - 1 \mid v \in V(T)\}$ and the *tree-width* of a graph $G$ is the minimum over the $\mathrm{width}(T, B)$, where $(T, B)$ is a tree decomposition of $G$.

Let $r$ be a positive integer. An *$r$-approximate linkage* in a graph $G$ is a family $L$ of paths with distinct endpoints in $G$ such that for every $r + 1$ distinct paths $P_1, P_2, \ldots, P_{r+1}$ in $L$, it holds that $\bigcap_{i \in [r+1]} V(P_i) = \emptyset$. We call these paths the *components* of the linkage. Let $(\alpha_1, \alpha_2, \ldots, \alpha_k)$ and $(\beta_1, \beta_2, \ldots, \beta_k)$ be elements of $V(G)^k$. We say that an $r$-approximate linkage $L$, consisting of the paths $P_1, P_2, \ldots, P_k$, *links* $(\alpha_1, \alpha_2, \ldots, \alpha_k)$ and $(\beta_1, \beta_2, \ldots, \beta_k)$ if $P_i$ is a path with endpoints $\alpha_i$ and $\beta_i$, for every $i \in [k]$. The *order* of such linkage is $k$. We call an $r$-approximate linkage of order $k$, *$r$-approximate $k$-linkage*. Two $r$-approximate $k$-linkages $L$ and $L'$ are *equivalent* if they have the same order and for every component $P$ of $L$ there exists a component $P'$ of $L'$ with the same endpoints. An $r$-approximate linkage $L$ of a graph $G$ is called *unique* if for every equivalent linkage $L'$ of $L$, $V(L) = V(L')$. When $r = 1$, such a family of paths is called *linkage*. Finally, a linkage $L$ in a graph $G$ is called *vital* if there is no other linkage in $G$ joining the same pairs of vertices.

In [32], N. Robertson and P. Seymour proved a theorem which is known as The Vital Linkage Theorem. This theorem provides an upper bound for the tree-width of a graph $G$ that contains a vital $k$-linkage $L$ such that $V(L) = V(G)$, where the bound depends only on $k$. A stronger statement of the Vital Linkage Theorem was recently proved by K. Kawarabayashi and P. Wollan [23], where instead of asking for the linkage to be vital, it asks for it to be unique. Notice here that a vital linkage is also unique. As in some of our proofs (for example, the proof of Lemma 5) we deal with unique but not necessarily vital linkages we make use of the Vital Linkage Theorem in its latter form which is stated below.

**Theorem 1** (The Unique Linkage Theorem [23, 32]). *There exists a computable function $w : \mathbb{N} \to \mathbb{N}$ such that the following holds. Let $L$ be a (1-approximate) $k$-linkage in $G$ with $V(L) = V(G)$. If $L$ is unique then $\mathbf{tw}(G) \leq w(k)$.*

## 2.3 Monadic Second Order Logic

We now recall some definitions from Monadic Second Order Logic (MSO). An extended introduction to Logic can be found in [16, 25].

We call *signature* $\tau = \{R_1, \ldots, R_n\}$ any finite set of relation symbols $R_i$ of any (finite) arity denoted by $\mathrm{ar}(R_i)$. For the language of graphs $\mathcal{G}$ we consider the signature $\tau_{\mathcal{G}} = \{V, E, I\}$ where $V$ represents the set of vertices of a graph $G$, $E$ the set of edges, and $I = \{(v, e) \,|\, v \in e \text{ and } e \in E(G)\}$ the incidence relation.

A $\tau$-structure $\mathfrak{A} = (A, R_1^{\mathfrak{A}}, \ldots, R_n^{\mathfrak{A}})$ consists of a finite universe $A$, and the interpretation of the relation symbols $R_i$ of $\tau$ in $A$, that is, for every $i$, $R_i^{\mathfrak{A}}$ is a subset of $A^{\mathrm{ar}(R_i)}$.

In MSO formulas are defined recursively from atomic formulas, that is, expressions of the form $R_i(x_1, x_2, \ldots, x_{\mathrm{ar}(R_i)})$ or of the form $x = y$ where $x_j$, $j \leq \mathrm{ar}(R_i)$, $x$ and $y$ are variables, by using the Boolean connectives $\neg, \land, \lor, \to$, and existential or universal quantification over individual variables and sets of variables.

Notice that in the language of graphs the atomic formulas are of the form $V(u), E(e)$ and $I(u, e)$, where $u$ and $e$ are vertex and edge variables respectively. Furthermore, quantification takes place over vertex or edge variables or vertex-set or edge-set variables.

A *graph structure* $\mathfrak{G} = (V(G) \cup E(G), V^{\mathfrak{G}}, E^{\mathfrak{G}}, I^{\mathfrak{G}})$ is a $\tau_{\mathcal{G}}$-structure, which represents the graph $G = (V, E)$. From now on, we abuse notation by treating $G$ and $\mathfrak{G}$ equally.

A graph class $\mathcal{C}$ is *MSO-definable* if there exists an MSO formula $\phi_{\mathcal{C}}$ in the language of graphs such that $G \in \mathcal{C}$ if and only if $G \models \phi_{\mathcal{C}}$, that is, $\phi_{\mathcal{C}}$ is true in the graph $G$ ($G$ is a model of $\phi_{\mathcal{C}}$).

**Lemma 1.** *The class of graphs that contain a fixed graph $H$ as an immersion is MSO-definable by an MSO-formula $\phi_H$.*

*Proof.* Let $V(H) = \{v_1, v_2, \ldots, v_n\}$ and $E(H) = \{e_1, e_2, \ldots, e_m\}$. Let also $\phi_H$ be the following formula.

$$\phi_H := \exists E_1, E_2, \ldots, E_m \exists x_1, x_2, \ldots, x_n \Big[ (\bigwedge_{i \in [n]} V(x_i)) \land (\bigwedge_{j \in [m]} E_i \subseteq E) \land$$
$$(\bigwedge_{i \neq j} x_i \neq x_j) \land (\bigwedge_{p \neq q} E_p \cap E_q = \emptyset) \land (\bigwedge_{e_r = \{v_k, v_l\} \in E(H)} \mathrm{path}(x_k, x_l, E_r)) \Big],$$

where path$(x, y, Z)$ is the MSO formula stating that the edges in $Z$ form a path from $x$ to $y$. This can be done by saying that the set $Z$ of edges is connected and every vertex $v$ incident to an edge in $Z$ is either incident to exactly two edges of $Z$ or to exactly one edge with further condition that $v = x$ or $v = z$. Thus, path$(x, y, Z)$ can be expressed in MSO by the following formula.

$$[(x \neq y) \wedge \exists p, q(Z(p) \wedge Z(q) \wedge I(x,p) \wedge I(y,q) \wedge$$
$$\forall p' \in Z(I(x,p') \to p = p') \wedge \forall q' \in Z(I(y,q') \to q = q')) \wedge$$
$$\forall w(V(w) \wedge w \neq x \wedge w \neq y \wedge \exists q_1(Z(q_1) \wedge I(w,q_1)) \to$$
$$\exists q_2, q_3(Z(q_2) \wedge Z(q_3) \wedge q_2 \neq q_3 \wedge I(w,q_2) \wedge I(w,q_3))) \wedge$$
$$\forall p_1, p_2, p_3(Z(p_1) \wedge Z(p_2) \wedge Z(p_3) \wedge$$
$$\exists m(V(m) \wedge I(u,p_1) \wedge I(u,p_2) \wedge I(u,p_3)) \to \bigvee_{i \neq j}(p_i = p_j))]$$

It is easy to verify that $\phi_H$ is the desired formula. $\square$

We now state a theorem which plays a crucial role in the proof of our algorithm for the computation of immersion obstructions for general immersion-closed graph classes.

**Theorem 2** ( [3,9]). *For every positive integer $k$, it is decidable given an MSO-formula whether it is satisfied by a graph $G$ whose tree-width is upper bounded by $k$, if $G$ is given together with a tree-decomposition.*

In [1], I. Adler, M. Grohe and S. Kreutzer provide tools that allow us to use Theorem 2, when an upper bound on the tree-width of the obstructions is known and an MSO-description of the graph class can be computed, in order to compute the obstruction sets of minor-closed graph classes. We adapt their machinery to the immersion ordering and prove that the tree-width of the obstructions of immersion-closed graph classes is upper bounded by some function that only depends on the graph class. This provides a generic technique to construct immersion obstruction sets when the explicit value of the function is known. Then, by obtaining such a computable upper bound on the tree-width of the graphs in $\mathbf{obs}_{\leq_{im}}(\mathcal{C})$, where $\mathcal{C} = \mathcal{C}_1 \cup \mathcal{C}_2$ and $\mathcal{C}_1$, $\mathcal{C}_2$ are immersion-closed graph classes whose obstruction sets are given, we show that the set $\mathbf{obs}_{\leq_{im}}(\mathcal{C})$ can be effectively computed.

## 3 Computing Immersion Obstruction Sets

In this Section we prove the analogue of Lemma 2.2 in [1] (Lemma 2) and the analogue of Lemma 3.1 in [1] (Lemma 4) for the immersion ordering.

We first state the combinatorial Lemma of this Section.

**Lemma 2.** *There exists a computable function $f : \mathbb{N} \to \mathbb{N}$ such that the following holds. Let $H$ and $G$ be graphs such that $H \leq_{im} G$. If $G'$ is a minimal subgraph of $G$ with $H \leq_{im} G'$ then $\mathbf{tw}(G') \leq f(|E(H)|)$.*

The proof of Lemma 2 is omitted as a stronger statement will be proved later on (Lemma 7). We continue by giving the necessary definitions in order to prove the analogue of Lemma 3.1 in [1] for the immersion ordering.

**Extension of MSO** For convenience, we consider the extension of the signature $\tau_\mathcal{G}$ to a signature $\tau_{ex}$ that pairs the representation of a graph $G$ with the representation of one of its tree-decompositions.

**Definition 2.** *If $G$ is a graph and $\mathcal{T} = (T, B)$ is a tree-decomposition of $G$, $\tau_{ex}$ is the signature that consists of the relation symbols $V, E, I$ of $\tau_\mathcal{G}$, and four more relation symbols $V_T, E_T, I_T$ and $B$.*
*A tree-dec expansion of $G$ and $\mathcal{T}$, is a $\tau_{ex}$-structure*

$$\mathfrak{G}_{ex} = (V(G) \cup E(G) \cup V(T) \cup E(T),$$
$$V^{\mathfrak{G}_{ex}}, E^{\mathfrak{G}_{ex}}, I^{\mathfrak{G}_{ex}}, V_T^{\mathfrak{G}_{ex}}, E_T^{\mathfrak{G}_{ex}}, I_T^{\mathfrak{G}_{ex}}, B^{\mathfrak{G}_{ex}})$$

*where $V_T^{\mathfrak{G}_{ex}} = V(T)$ represents the node set of $T$, $E_T^{\mathfrak{G}_{ex}} = E(T)$ the edge set of $T$, $I_T^{\mathfrak{G}_{ex}} = \{(v, e) \,|\, v \in e \cap V(T) \wedge e \in E(T)\}$ the incidence relation in $T$ and $B^{\mathfrak{G}_{ex}} = \{(t, v) \,|\, t \in V(T) \wedge v \in B_t \cap V(G)\}$.*

We denote by $\mathcal{C}_{\mathcal{T}_k}$ the class of tree-dec expansions consisting of a graph $G$ with $\mathbf{tw}(G) \leq k$, and a tree decomposition $(T, B)$ of $G$ of width$(T, B) \leq k$.

**Lemma 3** ( [1]).
  1. *Let $G$ be a graph and $(T, B)$ a tree decomposition of it with width$(T, B) \leq k$. Then, the tree-width of the tree-dec expansion of $G$ is at most $k + 2$.*
  2. *There is an MSO-sentence $\phi_{\mathcal{C}_{\mathcal{T}_k}}$ such that for every $\tau_{ex}$-structure $\mathfrak{G}$, $\mathfrak{G} \models \phi_{\mathcal{C}_{\mathcal{T}_k}}$ if and only if $\mathfrak{G} \in \mathcal{C}_{\mathcal{T}_k}$.*

A classic result [3] (see Theorem 2) states that we can decide, for every $k \geq 0$, if an MSO-formula is satisfied in a graph $G$ of $\mathbf{tw}(G) \leq k$. An immediate corollary of this result and Lemma 3 is the following.

**Corollary 1.** *We can decide, for every $k$, if an MSO-formula $\phi$ is satisfied in some $\mathfrak{G} \in \mathcal{C}_{\mathcal{T}_k}$.*

**Theorem 3** ( [1]). *For every $k \geq 0$, there is an MSO-sentence $\phi_{\mathcal{T}_k}$ such that for every tree-dec expansion $\mathfrak{G} \in \mathcal{C}_{\mathcal{T}_l}$ of $G$, for some $l \geq k$, it holds that $\mathfrak{G} \models \phi_{\mathcal{T}_k}$ if and only if $\mathbf{tw}(G) = k$.*

**Definition 3.** *A graph class $\mathcal{C}$ is layer-wise MSO-definable, if for every $k \in \mathbb{N}$ we can compute an MSO-formula $\phi_k$ such that $G \in \mathcal{C} \wedge \mathbf{tw}(G) \leq k$ if and only if $\mathfrak{G} \models \phi_k$, where $\mathfrak{G} \in \mathcal{C}_{\mathcal{T}_k}$ is the tree-dec expansion of $G$.*

**Definition 4.** *Let $\mathcal{C}$ be an immersion-closed graph class. The* **width** *of $\mathcal{C}$,* **width**$(\mathcal{C})$ *is the minimum positive integer $k$ such that for every graph $G \notin \mathcal{C}$ there is a graph $G' \subseteq G$ with $G' \notin \mathcal{C}$ and $\mathbf{tw}(G') \le k$.*

Note that Lemma 2 ensures that the width of an immersion-closed graph class is well-defined.

**Observation 1.** *If $\mathcal{C}_1$ and $\mathcal{C}_2$ are immersion-closed graph classes then the following hold.*

1. *For every graph $G \notin \mathcal{C}_1 \cup \mathcal{C}_2$, there exists a graph $G' \subseteq G$ such that $G' \notin \mathcal{C}_1 \cup \mathcal{C}_2$ and $\mathbf{tw}(G') \le \max\{r(|E(H)|, |E(J)|) \mid H \in \mathbf{obs}_{\le_{im}}(\mathcal{C}_1), J \in \mathbf{obs}_{\le_{im}}(\mathcal{C}_2)\}$, where $r$ is the function of Lemma 7 and thus,*

2. *For every graph $G \notin \mathcal{C}_1$, there exists a graph $G' \subseteq G$ such that $G' \notin \mathcal{C}_1$ and $\mathbf{tw}(G') \le \max\{f(|E(H)|) \mid H \in \mathbf{obs}_{\le_{im}}(\mathcal{C}_1)\}$, where $f$ is the function of Lemma 2.*

Finally, we state the analogue of Lemma 3.1 in [1] for the immersion ordering.

**Lemma 4.** *There exists an algorithm that, given an upper bound $l \ge 0$ on the width of a layer-wise MSO-definable class $\mathcal{C}$, and a computable function $f : \mathbb{N} \to MSO$ such that for every positive integer $k$, $f(k) = \phi_k$, where $\phi_k$ is the MSO-formula defining $\mathcal{C} \cap \mathcal{T}_k$, it computes $\mathbf{obs}_{\le_{im}}(\mathcal{C})$.*

*Proof.* In order to prove the Lemma it is enough to prove the following.

*Claim* 1. For any finite family of graphs $\mathcal{F} = \{F_1, \ldots, F_n\}$, it is decidable whether the following two following conditions are unsatisfiable for a given graph $G$.

1. $G \in C$ and there exists an $F \in \mathcal{F}$ such that $F \le_{im} G$.

2. $G \notin \mathcal{C}$ and for every $F \in \mathcal{F}$, $F \not\le_{im} G$.

To see that the above Claim is enough, first notice that if $\mathcal{F}$ is a finite family of graphs for which the formulas $\chi$ and $\psi$ are unsatisfiable then $\mathcal{F}$ is a forbidden immersion characterization of $\mathcal{C}$, that is, a graph $G$ belongs to $\mathcal{C}$ if and only if it does not contain any of the graphs in $\mathcal{F}$ as an immersion. By definition, $\mathbf{obs}_{\le_{im}}(\mathcal{C})$ is the minimum such family according to the relation $\le$ defined in Section 2. Thus, if Claim 1 holds, we can find the set $\mathbf{obs}_{\le_{im}}(\mathcal{C})$ by enumerating, according to $\le$, all the finite families of graphs $\mathcal{F}$ and deciding, for each one of them, if the formulas $\chi$ and $\psi$ are unsatisfiable.

*Proof of Claim 1.* Let $G$ be a graph in $\mathcal{C}$ such that $F \le_{im} G$, for some $F \in \mathcal{F}$. Lemma 2 implies that there exists a graph $G' \subseteq G$ such that $\mathbf{tw}(G') \le f(|E(F)|)$ and $F \le_{im} G'$, where $f$ is the function of Lemma 2. Observe that $G' \in \mathcal{C}$. Thus, $\chi$ is satisfiable if and only if there exists a graph in $\mathcal{C}$, whose tree-width is bounded from $\max\{f(|E(F)|) : F \in \mathcal{F}\}$, that satisfies it, where $f$ is the computable function of Lemma 2. Let $\phi_\mathcal{C}$ be the formula defining $\mathcal{C} \cap \mathcal{T}_k$ in $\mathcal{C}_{\mathcal{T}_k}$, and $\phi_\mathcal{F} \equiv \bigvee_{F \in \mathcal{F}} \phi_F$, where $\phi_F$ is the formula from Lemma 1 and $k = \max\{f(|E(F)|) : F \in \mathcal{F}\}$. Notice that there exists some graph $G \in \mathcal{C}$ that models $\phi_\mathcal{F}$ if and only if $\phi_\mathcal{C} \wedge \phi_\mathcal{F}$ is satisfiable for some $G' \in \mathcal{C}_{\mathcal{T}_k}$. From Corollary 1, this is decidable.

Let $G \notin \mathcal{C}$ be a graph such that $F \not\leq_{im} G$, for every $F \in \mathcal{F}$. Recall that the width of a graph class $\mathcal{C}$ is the minimum positive integer $k$ such that for every graph $G \notin \mathcal{C}$ there is a $G' \subseteq G$ with $G' \notin \mathcal{C}$ and $\mathbf{tw}(G') \leq k$. Thus, $G$ contains a subgraph $G'$ with tree-width at most $w$ such that $G' \notin \mathcal{C}$, where $w$ is computable by Lemma 2. Observe that $F \not\leq_{im} G'$, for every $F \in \mathcal{F}$. If $\phi'_\mathcal{C}$ is the MSO-sentence defining $\mathcal{C} \cap \mathcal{T}_w$ (given by the hypothesis), then there exists a graph $G \notin \mathcal{C}$ such that $F \not\leq_{im} G$, for every $F \in \mathcal{F}$ if and only if $\neg \phi'_\mathcal{C} \wedge \neg \phi_\mathcal{F}$ is satisfiable in $\mathcal{C}_{\mathcal{T}_w}$. The decidability of whether $\neg \phi'_\mathcal{C} \wedge \neg \phi_\mathcal{F}$ is satisfiable in $\mathcal{C}_{\mathcal{T}_w}$ follows, again, from Corollary 1. □

As Claim 1 holds, the lemma follows. □

**Corollary 2.** *There is an algorithm that given an MSO formula $\phi$ and $k \in \mathbb{N}$, so that $\phi$ defines an immersion closed-graph class $\mathcal{C}$ of width at most $k$, computes the obstruction set of $\mathcal{C}$.*

We would like to remark here that while Lemma 4 provides an algorithm for computing the obstruction set of any immersion-closed graph class $\mathcal{C}$, given that the conditions stated are satisfied, this result is generic and there is no uniform way for computing either an upper bound on the width of $\mathcal{C}$ or an MSO-description of $\mathcal{C}$.

In the next section, by proving some combinatorial lemmata, we are able to conclude that if $\mathcal{C}_1$ and $\mathcal{C}_2$ are two immersion-closed graph classes whose obstruction sets are known then the set $\mathbf{obs}_{\leq_{im}}(\mathcal{C}_1 \cup \mathcal{C}_2)$ is computable.

## 4 Tree-width Bounds for the Obstructions

In this section, we give an upper bound on the immersion obstruction set of the graph class $\mathcal{C}_1 \cup \mathcal{C}_2$ where $\mathcal{C}_1$ and $\mathcal{C}_2$ are immersion-closed graph classes, given that their obstruction sets are known. In order to do this, we first prove a generalization of the Unique Linkage Theorem. Then we introduce the notion of an $r$-approximate edge-linkage and work on the minimal graphs not belonging to $\mathcal{C}_1 \cup \mathcal{C}_2$.

Finally, as it is trivial to compute an MSO-description of $\mathcal{C}_1 \cup \mathcal{C}_2$ when we are given the sets $\mathbf{obs}_{\leq_{im}}(\mathcal{C}_1)$ and $\mathbf{obs}_{\leq_{im}}(\mathcal{C}_2)$, we show that the obstruction set of $\mathcal{C}_1 \cup \mathcal{C}_2$ is computable.

**Lemma 5.** *There exists a computable function $f : \mathbb{N} \to \mathbb{N}$ such that the following holds. Let $G$ be a graph that contains a 2-approximate $k$-linkage $\tilde{L}$ such that $V(\tilde{L}) = V(G)$. If $\tilde{L}$ is unique, then $\mathbf{tw}(G) \leq f(k)$.*

*Proof.* Let $G$ be a graph that contains a unique 2-approximate $k$-linkage $\tilde{L}$ with $V(\tilde{L}) = V(G)$ that links $A = (\alpha_1, \alpha_2, \ldots, \alpha_k)$ and $B = (\beta_1, \beta_2, \ldots, \beta_k)$ in $G$. Denote by $T$ the set $A \cup B$ and consider the graph $G^b$ with

$$\begin{aligned} V(G^b) &= V((G \setminus T) \times K_2) \cup T \\ E(G^b) &= E((G \setminus T) \times K_2) \cup \{\{t,t'\} \mid t,t' \in T \wedge \{t,t'\} \in E(G)\} \\ &\quad \cup \{\{t,(v,x)\} \mid t \in T \wedge x \in V(K_2) \wedge v \in V(G) \wedge \{t,v\} \in E(G)\}, \end{aligned}$$

where $V(K_2) = \{1, 2\}$. It is easy to see that $G^b$ contains a $k$-linkage that links $A$ and $B$. Let $G'$ be a minimal induced subgraph of $G^b$ that contains a $k$-linkage $L'$ that links $A$ and $B$. From Theorem 1, it follows that

$$\mathbf{tw}(G') \leq w(k). \tag{1}$$

From now on we work towards proving that $G \leq_m G'$. In order to achieve this, we prove the following two claims for $G'$.

*Claim* 2. If $L'$ is a $k$-linkage in $G'$ that links $A$ and $B$ then for every vertex $v \in V(G) \setminus T$ no path of $L'$ contains both $(v, 1)$ and $(v, 2)$.

*Proof.* Towards a contradiction, assume that for some vertex $v \in V(G) \setminus T$, there exists a $(t,t')$-path $P$ of $L'$ that contains both $(v, 1)$ and $(v, 2)$. Without loss of generality, assume also that $(v, 1)$ appears before $(v, 2)$ in $P$. Let $y$ be the successor of $(v, 2)$ in $P$ and notice that $y \neq (v, 1)$. From the definition of $G^b$ and the fact that $G'$ is an induced subgraph of $G^b$, $\{y, (v, 1)\} \in E(G') \setminus E(L')$. By replacing the subpath of $P$ from $(v, 1)$ to $y$ with the edge $\{(v, 1), y\}$, we obtain a linkage in $G' \setminus (v, 2)$ that links $A$ and $B$. This contradicts to the minimality of $G'$. □

*Claim* 3. If $L'$ is a $k$-linkage in $G'$ that links $A$ and $B$ then for every vertex $v \in V(G) \setminus T$, $V(L') \cap \{(v,1),(v,2)\} \neq \emptyset$.

*Proof.* Assume, in contrary, that there exists a linkage $L'$ in $G'$ and a vertex $x \in V(G) \setminus T$ such that $L'$ links $A$ and $B$ and $V(L') \cap \{(x,1),(x,2)\} = \emptyset$. Claim 2 ensures that, after contracting the edges $\{(v,1),(v,2)\}$, $v \in V(G) \setminus T$ (whenever they exist), the corresponding paths compose a 2-approximate $k$-linkage $\tilde{L}'$ of $G \setminus \{x\}$ that links $A$ and $B$. This is a contradiction to the assumption that $\tilde{L}$ is unique. Thus, the claim holds. □

Recall that $T \subseteq V(G')$ and that $G'$ is an induced subgraph of $G^b$. Claim 3 implies that we may obtain $G$ from $G'$ by contracting the edges $\{(v,1),(v,2)\}$ for every $v \in V(G) \setminus T$ (whenever they exist). As $G \leq_m G'$, from (1), it follows that, $\mathbf{tw}(G) \leq w(k)$. □

We remark that, the previous lemma holds for any graph $G$ that contains an $r$-approximate $k$-linkage. This can be seen by substituting $(G \setminus T) \times K_2$ with $(G \setminus T) \times K_r$ in its proof.

We now state a lemma that provides the upper bound of a graph $G$, given the upper bound of its line graph $L(G)$.

**Lemma 6.** *If $G$ is a graph and $k$ is a positive integer with $\mathbf{tw}(L(G)) \leq k$ then $\mathbf{tw}(G) \leq 2k+1$.*

*Proof.* Suppose that $G$ is graph such that $L(G)$ admits a tree decomposition of width at most $k$ and recall that every vertex of $L(G)$ corresponds to an edge of $G$. We construct a tree decomposition $\mathcal{T}$ of $G$ from a tree decomposition $\mathcal{T}_L$ of $L(G)$ by replacing in each bag of $\mathcal{T}_L$ every vertex of $L(G)$ by the endpoints of the corresponding edge in $G$. It is easy to verify that this is a tree decomposition of $G$. Therefore, $\mathbf{tw}(G) \leq 2k+1$. □

Before we proceed to the next lemma, we need to introduce the notion of an $r$-approximate $k$-edge-linkage in a graph. Similarly to the notion of an $r$-approximate linkage, an *$r$-approximate edge-linkage* in a graph $G$ is a family of paths $E$ in $G$ such that for every $r+1$ distinct paths $P_1, P_2, \ldots, P_{r+1}$ in $E$, it holds that $\cap_{i \in [r+1]} E(P_i) = \emptyset$. We call these paths the *components* of the edge-linkage. Let $(\alpha_1, \alpha_2, \ldots, \alpha_k)$ and $(\beta_1, \beta_2, \ldots, \beta_k)$ be elements of $V(G)^k$. We say that an $r$-approximate edge-linkage $E$, consisting of the paths $P_1, P_2, \ldots, P_k$, *links* $(\alpha_1, \alpha_2, \ldots, \alpha_k)$ and $(\beta_1, \beta_2, \ldots, \beta_k)$ if $P_i$ is a path with endpoints $\alpha_i$ and $\beta_i$, for every $i \in [k]$. The *order* of $E$ is $k$. We call an $r$-approximate edge-linkage of order $k$, $r$-approximate $k$-edge-linkage. When $r=1$, we call such a family of paths, an *edge-linkage*.

**Lemma 7.** *There exists a computable function $r$ such that the following holds. Let $G_1, G_2$ and $G$ be graphs such that $G_i \leq_{im} G$, $i = 1, 2$. If $G'$ is a minimal subgraph of $G$ where $G_i \leq_{im} G'$, $i = 1, 2$, then $\mathbf{tw}(G') \leq r(|E(G_1)|, |E(G_2)|)$.*

*Proof.* Let $G'$ be a minimal subgraph of $G$ such that $G_i \leq_{im} G'$, $i = 1, 2$. Recall that the edges of $G_i$ compose a $k_i$-edge-linkage $E_i$ in $G$, where $k_i = |E(G_i)|$, $i = 1, 2$. Furthermore, observe that the paths of $E_1$ and $E_2$ constitute a 2-approximate $k$-edge-linkage $E$ of $G$, where $k = k_1 + k_2$. Indeed, notice that in contrary to linkages, we do not require the paths that are forming edge-linkages to have different endpoints. The minimality of $G'$ implies that $\bigcup \{P \mid P \in E\} = G'$. Denote by $A = (v_{i_1}, v_{i_2}, \ldots, v_{i_k})$ and $B = (v_{j_1}, v_{j_2}, \ldots, v_{j_k})$ the vertex sets that are edge-linked by $E$ in $G'$ and let $\widehat{G}$ be the graph with

$$\begin{aligned} V(\widehat{G}) &= V(G') \cup \{u_{i_q} \mid q \in [k]\} \cup \{u_{j_q} \mid q \in [k]\}, \\ E(\widehat{G}) &= E(G') \cup \{t_{i_q} \mid q \in [k]\} \cup \{t_{j_q} \mid q \in [k]\}, \end{aligned}$$

where the vertices $u_{i_q}$ and $u_{j_q}$, $q \in [k]$ are new, $t_{i_q} = \{u_{i_q}, v_{i_q}\}$, $q \in [k]$ and $t_{j_q} = \{u_{j_q}, v_{j_q}\}$, $q \in [k]$.

Consider the line graph of $\widehat{G}$, $L(\widehat{G})$, and notice that $E$ corresponds to a 2-approximate $k$-linkage $L$ from $A_L$ to $B_L$ in $L(\widehat{G})$, where $A_L = (t_{i_1}, t_{i_2}, \ldots, t_{i_k})$ and $B_L = (t_{j_1}, t_{j_2}, \ldots, t_{j_k})$. This is true as, from the construction of $\widehat{G}$, all the vertices in $A_L$ and $B_L$ are distinct. The minimality of $G'$ yields that $V(L) = V(L(\widehat{G}))$ and implies that $L$ is unique. From Lemma 5, we obtain that $\mathbf{tw}(L(\widehat{G})) \leq f(k)$. Therefore, from Lemma 6, we get that $\mathbf{tw}(\widehat{G}) \leq p(f(k))$, where $p$ is the function of Lemma 6. Finally, as $G' \subseteq \widehat{G}$, $\mathbf{tw}(G') \leq r(k_1, k_2)$, where $r(k_1, k_2) = p(f(k_1 + k_2))$. □

Notice that Lemma 2 follows from Lemma 7 when we set $G_2$ to be the empty graph. Finally, we show that given two immersion-closed graph classes $\mathcal{C}_1$ and $\mathcal{C}_2$ the immersion-closed graph class $\mathcal{C}_1 \cup \mathcal{C}_2$ is layer-wise MSO-definable.

**Observation 2.** *Let $\mathcal{C}_1$ and $\mathcal{C}_2$ be immersion-closed graph classes, then $\mathcal{C} = \mathcal{C}_1 \cup \mathcal{C}_2$ is a layer-wise MSO-definable class defined, for every $k \geq 0$, by the formula*

$$\phi_k \equiv \left( \left( \bigwedge_{G \in \mathbf{obs}_{\leq_{im}}(\mathcal{C}_1)} \neg \phi_G \right) \vee \left( \bigwedge_{H \in \mathbf{obs}_{\leq_{im}}(\mathcal{C}_2)} \neg \phi_H \right) \right) \wedge \phi_{\mathcal{T}_k}$$

*where $\phi_G$ and $\phi_H$ are the formulas described in Lemma 1, and $\phi_{\mathcal{T}_k}$ the formula of Theorem 3 .*

We are now able to prove our main result.

**Theorem 4.** *Let $\mathcal{C}_1$ and $\mathcal{C}_2$ be two immersion-closed graph classes. If the sets $\mathbf{obs}_{\leq_{im}}(\mathcal{C}_1)$ and $\mathbf{obs}_{\leq_{im}}(\mathcal{C}_2)$ are given, then the set $\mathbf{obs}_{\leq_{im}}(\mathcal{C}_1 \cup \mathcal{C}_2)$ is computable.*

*Proof.* Observation 2, provides us with an MSO-description of the immersion-closed graph class $\mathcal{C}_1 \cup \mathcal{C}_2$, and Lemma 7 gives us an upper bound on the width of $\mathcal{C}_1 \cup \mathcal{C}_2$. Therefore, Lemma 4 is applicable. □

## 5   Conclusions and further work

In this paper, we further the study on the constructibility of obstruction sets for immersion-closed graph classes. In particular, we provide an upper bound on the tree-width of the obstructions of a graph class $\mathcal{C}$, which is the union of two immersion-closed graph classes $\mathcal{C}_1$ and $\mathcal{C}_2$ with $\mathbf{obs}_{\leq_{im}}(\mathcal{C}_1)$ and $\mathbf{obs}_{\leq_{im}}(\mathcal{C}_2)$ given. Then, using that result, we prove that $\mathbf{obs}_{\leq_{im}}(\mathcal{C})$ is computable.

In [33], N. Robertson and P. Seymour claimed that the class of graphs is also well-quasi-ordered under the strong immersion ordering. However, a full proof of this result has not appeared so far. We remark that the combinatorial results of this paper, that is, the upper bounds on the tree-width of the obstructions, also hold for the strong immersion ordering. Thus, if the claim of N. Robertson and P. Seymour holds, the obstruction set of the union of two strongly immersion-closed graph classes, whose obstruction sets are given, can be effectively computed.

Finally, it was proven by B. Courcelle, R. Downey and M. Fellows [10] that the obstruction set of a minor-closed graph class $\mathcal{C}$ cannot be computed by an algorithm whose input is a description of $\mathcal{C}$ as an MSO-sentence. The computability of the obstruction set of an immersion-closed graph class $\mathcal{C}$, given solely an MSO description of $\mathcal{C}$, remains an open problem.

**Acknowledgements.** We would like to thank Prof. Dimitrios M. Thilikos for suggesting this problem to us. We also wish to thank him, and the anonymous referees of a preliminary version of this paper [21], for helpful comments and remarks that improved the presentation of our results.